\documentclass[12pt,fleqn]{article}
\usepackage[dvips]{graphicx}
\begin{document}

\baselineskip 20pt

\begin{flushright}
\begin{tabular}{l}
NEAP-53, April 1997 \\
hep-ph/9705203
\end{tabular}
\end{flushright}

\vspace*{1.5cm}

\begin{center}

{\Large{\bf Flavor nonconservation and $ CP $ violation}}

{\Large{\bf from quark mixings with singlet quarks}}

\vspace{1cm}

{\large{Isao Kakebe $ {}^1 $,~ Katsuji Yamamoto $ {}^2 $}}

\vspace{1cm}

{\it Department of Nuclear Engineering, Kyoto University,
Kyoto 606-01, Japan}

\end{center}

\vspace{1cm}

{\bf abstract}

\bigskip
Flavor nonconserving and $ CP $ violating effects of the quark mixings
are investigated in electroweak models
incorporating singlet quarks.  Especially, the $ D^0 $-$ {\bar D}^0 $
mixing and the neutron electric dipole moment, which are mediated
by the up-type quark couplings to the neutral Higgs fields,
are examined in detail for reasonable ranges of the quark mixings
and the singlet Higgs mass scale.
These neutral Higgs contributions are found to be comparable to
or smaller than the experimental bounds
even for the case where the singlet Higgs mass scale is
of the order of the electroweak scale
and a significant mixing is present
between the top quark and the singlet quarks.

\vspace*{1cm}
\begin{flushleft}



\vspace*{1cm}

$ {}^1 $ e-mail address: kakebe@b1.nucleng.kyoto-u.ac.jp

$ {}^2 $ e-mail address: yamamoto@b1.nucleng.kyoto-u.ac.jp

\end{flushleft}

\newpage
Some extensions of the minimal standard model may be motivated
in various points of view.
For instance, in the electroweak baryogenesis,
the $ CP $ asymmetry induced by the conventional phase
in the Cabibbo-Kobayashi-Maskawa (CKM) matrix
is far too small to account for the observed baryon to entropy ratio,
and the electroweak phase transition should be at most of weakly first order
consistently with the experimental bound on the Higgs particle mass
\cite{baryogenesis}.
In this article, among various possible extensions, we investigate
electroweak models incorporating $ {\rm SU(2)}_W \times {\rm U(1)}_Y $
singlet quarks with electric charges $ 2/3 $ and/or $ - 1/3 $
\cite{Br1,Ag1,B-M,q-Qmix,Br2,L-S,Br3,Rb,KY,Ag2}.
These sorts of models have some novel features arising from the mixings
between the ordinary quarks ($ q $) and the singlet quarks ($ Q $):
The CKM unitarity in the ordinary quark sector is violated,
and the flavor changing neutral currents (FCNC's) are present
at the tree-level.  Furthermore, the $ q $-$ Q $ mixings involving
new $ CP $ violating sources are expected to make
important contributions to the electroweak baryogenesis
\cite{tUbaryogenesis}.
A singlet Higgs field $ S $ introduced to provide the singlet quark mass
terms and $ q $-$ Q $ mixing terms is even preferable for realizing
a strong enough first order electroweak phase transition
\cite{baryogenesis}.
It should also be mentioned, as investigated in the earlier
literature
\cite{Br2},
that the complex vacuum expectation value (vev)
$ \langle S \rangle $ of the singlet Higgs field induced
by the spontaneous $ CP $ violation leads to a non-vanishing CKM phase.

The quark mixings with singlet quarks are,
on the other hand, subject to the constraints coming
from various flavor nonconserving and $ CP $ violating processes
\cite{Br1,Ag1,B-M,q-Qmix,Br2,Br3,KY,Ag2}.
In particular, it is claimed in
\cite{Br2}
with a simple calculation for the case of $ d $-$ D $ mixings
that a rather stringent bound,
$ v_S \equiv | \langle S \rangle |
{\mbox{~$ > $ \hspace{-1.05em}{\raisebox{-0.75ex}{$ \sim $}}~}}
{\rm several~TeV} $, on the singlet Higgs mass scale is imposed
from the one-loop neutral Higgs contributions
to the neutron electric dipole moment (NEDM).
(The NEDM in electroweak models with singlet quarks
is also considered in \cite{Ag1,B-M}.)
The singlet Higgs field with mass scale in TeV region,
however, seems to be unfavorable for the electroweak baryogenesis.
Considering this apparently controversial situation, we examine
in detail such phenomenological implications of the electroweak models
with singlet quarks.
We mainly describe the case of singlet $ U $ quarks
with electric charge 2/3 in the following,
keeping in mind the possibility of significant $ t $-$ U $ mixing
for the electroweak baryogenesis.
The analyses are extended readily to the case of singlet $ D $ quarks
with electric charge $ -1/3 $, and similar results are obtained
for both cases.
In practice, diagonalization of the quark mass matrix is made numerically,
in order to calculate precisely the relevant quark couplings
to the neutral Higgs fields involving flavor nonconservation
and $ CP $ violation.  Then, for reasonable ranges of the model parameters,
systematic analyses are performed on the neutral Higgs contributions
to the $ D^0 $-$ {\bar D}^0 $ mixing and the NEDM.
In contrast to the earlier expectation
\cite{Br2},
they will turn out to be comparable to or smaller
than the experimental bounds
even for the case where the singlet Higgs mass scale is of the order
of the electroweak scale and a significant $ t $-$ U $ mixing is present.

The Yukawa couplings relevant for the up-type quark sector are given by
\begin{equation}
{\cal L}_{\rm Yukawa}
= - u^c_0 \lambda_u q_0 H - u^c_0 ( f_U S + f_U^\prime S^\dagger ) U_0
- U^c_0 ( \lambda_U S + \lambda_U^\prime S^\dagger ) U_0 ~+~{\rm h.c.}
\label{eqn:LYukawa}
\end{equation}
with the two-component Weyl fields (the generation indices and the factors
representing the Lorentz covariance are omitted for simplicity).
Here $ q_0 = ( u_0 , V_0 d_0 ) $ represents the quark doublets
with a unitary matrix $ V_0 $,
and $ H = ( H^0 , H^+ ) $ is the electroweak Higgs doublet.
A suitable redefinition among the $ u^c_0 $ and $ U^c_0 $ fields
with the same quantum numbers has been made
to eliminate the $ U^c_0 q_0 H $ couplings without loss of generality.
Then, the Yukawa coupling matrix $ \lambda_u $ has been
made diagonal by using unitary transformations
among the ordinary quark fields.
In this basis, by turning off the $ u $-$ U $ mixings
with $ f_U , f_U^\prime \rightarrow 0 $,
$ u_0 $ and $ d_0 $ are reduced to the mass eigenstates,
and $ V_0 $ is identified with the CKM matrix.
The actual CKM matrix is slightly modified due to the $ u $-$ U $ mixings
(and possibly the $ d $-$ D $ mixings), as shown explicitly later.
The Higgs fields develop vev's,
\begin{equation}
\langle H^0 \rangle = \frac{v}{\sqrt 2} ~,~~
\langle S \rangle = {\rm e}^{i \phi_S} \frac{v_S}{\sqrt 2}~,
\label{eqn:vev}
\end{equation}
where $ \langle S \rangle $ may acquire a nonvanishing phase $ \phi_S $
due to either spontaneous or explicit $ CP $ violation
originating in the Higgs sector.
The quark mass matrix is produced with these vev's as
\begin{equation}
{\cal M}_{\cal U} = \left( \begin{array}{cc}
M_u & \Delta_{u{\mbox{-}}U} \\ 0 & M_U \end{array} \right) ~,
\label{eqn:MUmatrix}
\end{equation}
where
\begin{equation}
M_u = \frac{\lambda_u v}{\sqrt 2}~,~~
\Delta_{u{\mbox{-}}U}
= ( f_U {\rm e}^{i \phi_S} + f_U^\prime {\rm e}^{-i \phi_S} )
\frac{v_S}{\sqrt 2} ~,~~
M_U = ( \lambda_U {\rm e}^{i \phi_S}
+ \lambda_U^\prime {\rm e}^{-i \phi_S} ) \frac{v_S}{\sqrt 2}~.
\label{eqn:MUsub}
\end{equation}

The quark mass matrix is diagonalized by unitary transformations
$ {\cal V}_{{\cal U}_{\rm L}} $ and $ {\cal V}_{{\cal U}_{\rm R}} $ as
\begin{equation}
{\cal V}_{{\cal U}_{\rm R}}^\dagger {\cal M}_{\cal U}
{\cal V}_{{\cal U}_{\rm L}}
= {\rm diag.} ( m_u , m_c , m_t , m_U ) ~.
\label{eqn:MUdia}
\end{equation}
(While the case with one singlet $ U $ quark is described hereafter
for simplicity of notation, the analyses are readily extended to the case
with more than one singlet $ U $ quarks, resulting analogous conclusions.)
The quark mass eigenstates are determined
in terms of the original states by
\begin{equation}
\left( \begin{array}{c} u \\ U \end{array} \right)
= {\cal V}_{{\cal U}_{\rm L}}^\dagger
\left( \begin{array}{c} u_0 \\ U_0 \end{array} \right) ~,~~
( u^c , U^c )
= ( u_0^c , U_0^c ){\cal V}_{{\cal U}_{\rm R}} ~.
\label{eqn:MUeigen}
\end{equation}
The unitary transformations for the quark mixings may be represented as
\begin{equation}
{\cal V}_{{\cal U}_\chi}
= \left( \begin{array}{cc}
V_{u_\chi} & \epsilon_{u_\chi} \\
{ } & { } \\
- \epsilon_{u_\chi}^{\prime \dagger} & V_{U_\chi}
\end{array} \right) ~~ [ \chi = {\rm L} , {\rm R} ] .
\label{eqn:Vchi}
\end{equation}
(Similar unitary transformations $ {\cal V}_{{\cal D}_\chi} $
are introduced if the $ d $-$ D $ mixings are also present.)
Then, the $ u $-$ U $ mixing submatrices are found in the leading orders
to be
\begin{equation}
( \epsilon_{u_{\rm L}} )_i \simeq ( \epsilon_{u_{\rm L}}^\prime )_i
\simeq ( M_u \Delta_{u{\mbox{-}}U} M_U^{-2} )_i
\sim ( m_{u_i} / m_U ) \epsilon_{u{\mbox{-}}U} ~,
\label{eqn:epsL}
\end{equation}
\begin{equation}
( \epsilon_{u_{\rm R}} )_i \simeq ( \epsilon_{u_{\rm R}}^\prime )_i
\simeq ( \Delta_{u{\mbox{-}}U} M_U^{-1} )_i \sim \epsilon_{u{\mbox{-}}U} ~.
\label{eqn:epsR}
\end{equation}
Here the parameter $ \epsilon_{u{\mbox{-}}U}
\sim ( | f_U | + | f_U^\prime | )/( | \lambda_U |
+ | \lambda_U^\prime | ) $ represents the mean magnitude
of the $ u $-$ U $ mixings, though there may be some generation dependence
more precisely, as seen later.
The relation between the masses and Yukawa couplings of the ordinary
up-type quarks is slightly modified by the $ u $-$ U $ mixings as
\begin{equation}
m_{u_i} \simeq \frac{\lambda_{u_i} v}{\sqrt 2} \left[ 1 - \frac{1}{2}
( \epsilon_{u_{\rm R}} \epsilon_{u_{\rm R}}^\dagger )_{ii} \right] ~.
\label{eqn:mui}
\end{equation}
The generalized CKM matrix for the up-type quarks including
the $ U $ quark is given by
\begin{equation}
\left( \begin{array}{cc}
V & \begin{array}{c} 0 \\ 0 \end{array}
\end{array} \right)
= {\cal V}_{{\cal U}_{\rm L}}^\dagger
\left( \begin{array}{cc}
V_0 & 0 \\ 0 & 0 \end{array} \right)
= \left( \begin{array}{cc}
V_{u_{\rm L}}^\dagger V_0 & 0 \\
\epsilon_{u_{\rm L}}^\dagger V_0 & 0 \end{array} \right) ~.
\label{eqn:VCKM}
\end{equation}
(The effect of possible $ d $-$ D $ mixings may be included
by multiplying $ {\cal V}_{{\cal D}_{\rm L}} $ from the right.)
Here the CKM unitarity within the ordinary quark sector
is violated slightly due to the $ u $-$ U $ mixings.

It should be noticed in eq.(\ref{eqn:epsL})
that the left-handed $ u $-$ U $ mixings are suppressed further
by the relevant $ u $-$ U $ mass ratios $ m_{u_i} / m_U $.
On the other hand, the right-handed $ u $-$ U $ mixings given
in eq.(\ref{eqn:epsR}) are actually ineffective by themselves,
since $ u^c $ and $ U^c $ fields have the same quantum numbers.
Hence, in models of this sort with the quark mass matrix of the form
given in eqs.(\ref{eqn:MUmatrix}) and (\ref{eqn:MUsub}),
the effects of the light ordinary quark mixings
with the singlet quarks appear to be rather small
with suppression factors of $ \epsilon_{u{\mbox{-}}U} $
and $ m_{u_i}/m_U $ under the natural relation
$ \lambda_{u_i} \sim m_{u_i}/v $.
In fact, the CKM unitarity violation within the ordinary quark sector
is found to arise at the order of
$ ( m_{u_i} m_{u_j} / m_U^2 ) \epsilon_{u{\mbox{-}}U}^2 $
\cite{Br1,Ag1,q-Qmix,Br2,KY,Ag2},
which is sufficiently below the experimental bounds
\cite{ParticleData}.
Contrary to this situation, it is in principle possible, for instance,
to take $ \lambda_{u_1} \gg m_u / v $ by making a fine-tuning
in eq.(\ref{eqn:mui})
with a significant mixing between the $ u (= u_1) $ quark
and the singlet $ U $ quark.
However, such a choice will not respect the quark mass hierarchy
in a natural sense; the smallness of $ m_u $ is no longer guaranteed
by a chiral symmetry appearing for $ \lambda_{u_1} \rightarrow 0 $.

The quark couplings to the neutral Higgs fields are extracted
from the Yukawa couplings (\ref{eqn:LYukawa}),
which involve flavor nonconservation and $ CP $ violation
induced by the $ u $-$ U $ mixings:
\begin{equation}
{\cal L}_{\rm Yukawa}^{\rm neutral}
= - \sum_{a = 0,1,2} {\cal U}^c \Lambda_a {\cal U} \phi_a ~+~ {\rm h.c.}~,
\label{eqn:LYneutral}
\end{equation}
where $ {\cal U} = ( u , c , t , U ) $
represents the quark mass eigenstates,
and $ \phi_0 $, $ \phi_1 $, $ \phi_2 $ are the mass eigenstates
of the neutral Higgs fields.
The original complex Higgs fields are decomposed as
$ H^0 = \langle H^0 \rangle + ( h_1 + i h_2 )/{\sqrt 2} $
and $ S = \langle S \rangle + ( s_1 + i s_2 )/{\sqrt 2} $
with real fields.
While the Nambu-Goldstone mode $ h_2 $ is absorbed by the $ Z $ gauge boson,
the remaining $ h_1 $, $ s_1 $, $ s_2 $ are combined
to form the mass eigenstates $ \phi_a $.
Then, the coupling matrices in eq.(\ref{eqn:LYneutral}) are given by
\begin{equation}
\Lambda_a = \frac{1}{\sqrt 2} {\cal V}_{{\cal U}_{\rm R}}^\dagger
( {\mbox{\sf O}}_{a0} \Lambda_u + {\mbox{\sf O}}_{a1} \Lambda_{U}^+
+ i {\mbox{\sf O}}_{a2} \Lambda_{U}^- ) {\cal V}_{{\cal U}_{\rm L}}
\label{eqn:Lam-a}
\end{equation}
with
\begin{equation}
\Lambda_u = \left( \begin{array}{cc}
\lambda_u & 0 \\
0 & 0
\end{array} \right) ~,~~
\Lambda_{U}^\pm = \left( \begin{array}{cc}
0 & f_U \pm f^\prime_U \\
0 & \lambda_U \pm \lambda^\prime_U
\end{array} \right) ~.
\label{eqn:Lam-uU}
\end{equation}
Here the Higgs mass eigenstates are expressed
with a suitable orthogonal matrix $ {\mbox{\sf O}} $ as
\begin{equation}
\left( \begin{array}{c} \phi_0 \\ \phi_1 \\ \phi_2 \end{array} \right)
= {\mbox{\sf O}} \left( \begin{array}{c} h_1 \\ s_1 \\
s_2 \end{array} \right) ~.
\label{eqn:phi-a}
\end{equation}
At present the Higgs masses $ m_{\phi_a} $ and the mixing matrix
$ {\mbox{\sf O}} $ should be regarded as free parameters
varying in certain reasonable ranges.

The quark mass matrix may be diagonalized perturbatively with respect
to the relevant couplings to determine the quark mixing matrices
$ {\cal V}_{{\cal U}_{\rm L}} $ and $ {\cal V}_{{\cal U}_{\rm R}} $.
Then, it is seen
from eqs.(\ref{eqn:Lam-a}) and (\ref{eqn:Lam-uU})
that the FCNC's of the ordinary up-type quarks coupled to
the neutral Higgs fields, in particular, have the following specific
generation dependence:
\begin{equation}
( \Lambda_a )_{ij}
\sim ( m_{u_j} / m_U ) \epsilon_{u{\mbox{-}}U}^2 ~( i \not= j )~.
\label{eqn:Lam-a-ij}
\end{equation}
This feature is actually confrimed by numerical calculations,
and is also valid for the case of $ d $-$ D $ mixings.
Since these FCNC's coupled to the neutral Higgs fields
are of the first order of the ordinary quark Yukawa couplings,
their contributions are expected to be significant
in certain flavor nonconserving and $ CP $ violating processes
such as the NEDM, $ D^0 $-$ {\bar D}^0 $,
$ K^0 $-$ {\bar K}^0 $, $ B^0 $-$ {\bar B}^0 $,
$ b \rightarrow s \gamma $, and so on
\cite{Br1,Ag1,B-M,q-Qmix,Br2,Br3,KY,Ag2}.
In fact it will be seen below that the neutral Higgs contributions
to the $ D^0 $-$ {\bar D}^0 $ mixing and the NEDM
become important for reasonable ranges of the model parameters.
In contrast to the FCNC's coupled to the neutral Higgs fields,
the $ q $-$ Q $ mixing effects on the $ Z $ gauge boson
couplings appear at the order of
$ ( m_{u_i} m_{u_j} / m_U^2 ) \epsilon_{u{\mbox{-}}U}^2 $
with the relation $ m_{u_i} \sim \lambda_{u_i}/v $,
which are related to the CKM unitarity violation
\cite{Br1,Ag1,q-Qmix,KY,Ag2}.
Since they are of the second order of the $ u $-$ U $ mass ratios,
the contributions of the FCNC's in gauge couplings are suppressed
much more, being sufficiently below the experimental bounds
\cite{ParticleData}
for the natural choices of the model parameters.
Detailed analyses on various flavor nonconserving
and $ CP $ violating effects coming from the quark mixings
with singlet quarks will be presented elsewhere,
which are, in particular, mediated
by the quark couplings to the neutral Higgs fields.
They would serve as signals for the new physics
beyond the minimal standard model.

The effective Hamiltonian relevant for the $ D^0 $-$ {\bar D}^0 $ mixing
is obtained with the quark couplings in eq. (\ref{eqn:Lam-a})
mediated by the neutral Higgs fields $ \phi_a $.
They are written with the four-component Dirac fields as
\begin{equation}
{\cal H}_{\phi}^{\Delta c = 2}
= \sum_a \frac{1}{m_{\phi_a}^2} \left[ {\bar c}
\left\{ ( \Gamma_a^S )_{21}
+ ( \Gamma_a^P )_{21} \gamma_5 \right\} u \right]^2 ~,
\label{eqn:H-phi}
\end{equation}
where
\begin{equation}
( \Gamma_a^S )_{21} = \frac{1}{2} \left[ ( \Lambda_a )_{12}^*
+ ( \Lambda_a )_{21} \right] ~,~~
( \Gamma_a^P )_{21} = \frac{1}{2} \left[ ( \Lambda_a )_{12}^*
- ( \Lambda_a )_{21} \right] ~.
\label{eqn:Gamma-SP}
\end{equation}
The $ D^0 $-$ {\bar D}^0 $ transition matrix element is calculated
with this effective Hamiltonian as follows
\cite{DDbar}:
\begin{eqnarray}
\langle {\bar D}^0 | {\cal H}_{\phi}^{\Delta c = 2} | D^0 \rangle
&=& \sum_a \frac{1}{6 m_{\phi_a}^2} \left[
\left\{ ( \Gamma_a^S )_{21}^2 - ( \Gamma_a^P )_{21}^2 \right\}
{\cal M}_A^0 \right. \nonumber \\
& { } & \left. ~~~~~~~+~ \left\{ -( \Gamma_a^S )_{21}^2
+ 11( \Gamma_a^P )_{21}^2 \right\} {\cal M}_P^0 \right] ~,
\label{eqn:DDbarmatrix}
\end{eqnarray}
\begin{equation}
{\cal M}_A^0 = f_D^2 m_D^2 ~,~~
{\cal M}_P^0 = \frac{f_D^2 m_D^4}{( m_u + m_c )^2} ~.
\label{eqn:M-AP}
\end{equation}
(In the present analysis, it is enough to use the vacuum insertion
approximation, by considering various ambiguities in choosing
the model parameter values.)
Then, the neutral Higgs contribution
to the neutral $ D $ meson mass difference is given by
\begin{equation}
\Delta m_D ( \phi ) = \left| \frac{1}{m_D} \langle {\bar D}^0 |
{\cal H}_{\phi}^{\Delta c = 2} | D^0 \rangle \right| ~.
\end{equation}
It is seen from eq.(\ref{eqn:Lam-a-ij}) that $ ( \Lambda_a )_{12} $
is, in particular, proportional to $ m_c / m_U $ rather than
$ m_u / m_U $, providing a significant contribution
to the $ D^0 $-$ {\bar D}^0 $ mixing.
The $ Z $ boson contribution to the $ D^0 $-$ {\bar D}^0 $ mixing
is investigated
in ref. \cite{Br3}
by considering a possible significant FCNC
between the $ c $ and $ u $ quarks.
In contrast to that analysis, the $ Z $ boson FCNC of $ u $ and $ c $
arises at the order of $ ( m_u m_c / m_U^2 ) \epsilon_{u{\mbox{-}}U}^2 $
with the quark mass matrix (\ref{eqn:MUmatrix})
respecting the relation $ \lambda_{u_i} \sim m_{u_i} / v $.
Hence its contribution to the $ D^0 $-$ {\bar D}^0 $ mixing
becomes rather small in the present case.

Important contributions are also provided to the NEDM
from the quark mixings with singlet quarks.
Here, the $ u $ quark EDM, which is one of the main components of the NEDM,
is induced by the $ u $-$ U $ mixings
in the one-loop diagrams involving the quark and neutral Higgs
intermediate states.
(The gauge couplings, on the other hand, do not contribute
to the NEDM at the one-loop level, which is due to the same situations
as in the minimal standard model
\cite{CPviolation,KY}.)
The total contribution of the neutral Higgs fields $ \phi_a $
to the $ u $ quark EDM is calculated by a formula
\begin{equation}
d_u ( \phi ) = - \frac{2 e}{3 (4 \pi )^2} \sum_a
\sum_{{\cal U}_K = u_i , U} \left\{
{\rm Im} \left [  ( \Lambda_a )_{1 K} ( \Lambda_a )_{K 1}
\right] \frac{m_{{\cal U}_K}}{m_{\phi_a}^2}
I ( m_{{\cal U}_K}^2 / m_{\phi_a}^2 ) \right\} ~,
\label{eqn:du-phi}
\end{equation}
where
\begin{equation}
I(X) = \frac{1}{(1 - X)^2} \left[ - \frac{3}{2} + \frac{1}{2} X
- \frac{1}{1 - X} \ln X \right] ~,
\label{eqn:I(X)}
\end{equation}
and $ m_{{\cal U}_K} = m_{u_i} , m_U $ represent the quark mass
eigenvalues.
(This form of $ I(X) $ may be modified for the intermediate
state of $ {\cal U}_K = u_1 $.  However, the contribution with
the $ u $ quark intermediate state is in any case negligible
due to the very small mass $ m_u $.)
These contributions to the $ u $ quark EDM in fact arise
at the first oder of the $ u $ quark mass $ m_u $
($ \lambda_{u_1} \sim m_u / v $).
This feature is understood by considering the limit
$ \lambda_{u_1} \rightarrow 0 $, where the $ u $ quark EDM is vanishing
due to a relevant chiral symmetry.
It is also noticed from eq.(\ref{eqn:Lam-a-ij})
that the top quark contributions with the couplings factors
$ ( \Lambda_a )_{13} ( \Lambda_a )_{31}
\sim ( m_u m_t / m_U^2 ) \epsilon_{u{\mbox{-}}U}^4 $
become important together with those of
the singlet $ U $ quark intermediate states.

We now make a detailed analysis of the $ u $-$ U $ mixing effects
on $ D^0 $-$ {\bar D}^0 $ mixing and the $ u $ quark EDM.
Numerical calculations are performed systematically in the following way:
First, by taking the model parameters in certain reasonable ranges,
the quark mass matrix $ {\cal M}_{\cal U} $ is diagonalized numerically
to obtain the quark mixing matrices
$ {\cal V}_{{\cal U}_{\rm L}} $ and $ {\cal V}_{{\cal U}_{\rm R}} $.
Then, the quark couplings $ \Lambda_a $ to the neutral Higgs fields
are determined with eqs. (\ref{eqn:Lam-a}) and (\ref{eqn:Lam-uU}). 
Finally, $ \Delta m_D ( \phi ) $ and $ d_u ( \phi ) $
are calculated by using the formulae
(\ref{eqn:H-phi}) -- (\ref{eqn:I(X)}).

Practically, the relevant model parameters are taken as follows:
The Yukawa couplings to the singlet Higgs field may be
parametrized by considering eqs.(\ref{eqn:MUsub}) and (\ref{eqn:epsR}) as
\begin{equation}
( f_U )_i = \kappa_i \epsilon_{u{\mbox{-}}U}
{\rm e}^{i \alpha_i} ( | M_U | / v_S ) ~,~~
( f_U^\prime )_i = \kappa^\prime_i \epsilon_{u{\mbox{-}}U}
{\rm e}^{i \alpha^\prime_i} ( | M_U | / v_S ) ~,
\label{eqn:Y-f}
\end{equation}
\begin{equation}
\lambda_U {\rm e}^{i \phi_S} + \lambda_U^\prime {\rm e}^{-i \phi_S}
\equiv | \lambda_U | {\rm e}^{i \beta}
+ | \lambda_U^\prime | {\rm e}^{i \beta^\prime}
= {\sqrt 2} ( M_U / v_S ) ~.
\label{eqn:Y-lam}
\end{equation}
The generation dependence of the $ u $-$ U $ couplings is taken
into account with the factors $ \kappa_i $ and $ \kappa^\prime_i $.
The $ \lambda_U $ and $ \lambda_U^\prime $ couplings
are taken under the condition in eq. (\ref{eqn:MUsub})
so as to reproduce the given value of $ | M_U | $
with varying $ v_S $, where $ | M_U | $ is approximately equal
to the $ U $ quark mass $ m_U $ up to the corrections
due to the $ u $-$ U $ mixings.
More specifically these parameters are taken in the ranges of
\begin{equation}
\epsilon_{u{\mbox{-}}U} \sim 0.1 ~,~~
0 \leq \kappa_i ,~ \kappa^\prime_i \leq 1 ~,~~
0 \leq \alpha_i ,~ \alpha^\prime_i < 2 \pi~,~~
\label{eqn:uUmix}
\end{equation}
\begin{equation}
| M_U | \simeq m_U \sim {\rm several} \times 100 {\rm GeV} ~,~~
0 \leq \beta ,~ \beta^\prime < 2 \pi ~.
\label{eqn:MUlam}
\end{equation}
Here the $ U $ quark with $ m_U \sim m_t $ is favored,
since significant contributions to the electroweak baryogenesis
are then expected to be obtained through the $ CP $
violating $ t $-$ U $ mixing
\cite{tUbaryogenesis}.
(It is, however, necessary to take at least $ m_U > m_t $,
since the $ U $ quark has a dominant decay mode
$ U \rightarrow b + W $ similarly to the top quark.)
The diagonal coupling matrix $ \lambda_u $ for the ordinary
up-type quarks is, on the other hand, chosen suitably
with the relation (\ref{eqn:mui})
so that the quark masses $ m_u $, $ m_c $ and $ m_t $
are reproduced within the experimentally determined ranges
\cite{ParticleData}.
(The corrections due to the $ u $-$ U $ mixings
in eq. (\ref{eqn:mui}) are actually at most of 10 \%
for $ \epsilon_{u{\mbox{-}}U} \sim 0.3 $.)
The decay constant of the $ D $ meson is taken to be
$ f_D = 0.3 {\rm GeV} $
\cite{ParticleData}.
The parameters concerning the Higgs fields are taken as
\begin{equation}
m_{\phi_0} \sim 100 {\rm GeV} ~,~~
m_{\phi_1} , m_{\phi_2} \sim v_S
{\mbox{~$ > $ \hspace{-1.05em}{\raisebox{-0.75ex}{$ \sim $}}~}}
100 {\rm GeV} ~,
\end{equation}
\begin{equation}
{\mbox{\sf O}}_{0 1} , {\mbox{\sf O}}_{0 2} \sim v / v_S ~,~~
{\mbox{\sf O}}_{12} \sim 1 ~,~~ 0 \leq \phi_S < 2 \pi ~.
\end{equation}
Here the mass of $ \phi_0 $ is fixed to be a somewhat smaller value
of 100 GeV or so, as suggested from the requirement
that the first order electroweak phase transition be strong enough.
The mixings between $ h_1 $ and $ s_1 $, $ s_2 $ are supposed
to scale as $ v/v_S $ in a viewpoint of naturalness,
since $ \phi_0 \approx h_1 $ (the standard neutral Higgs)
for the extreme case of $ v_S \gg v \simeq 246 {\rm GeV} $.

The resultant $ | d_u ( \phi ) | $ versus $ \Delta m_D ( \phi ) $
are shown in figs. 1 and 2 together with the experimental upper bounds
on the NEDM (dashed line)
and on $ \Delta m_D \equiv | m_{D^0_1} - m_{D^0_2} | $ (dotted line)
\cite{ParticleData}, where the relevant model parameters
are taken typically as
$ \epsilon_{u{\mbox{-}}U} = 0.3 $, $ M_U = 300~{\rm GeV} $ ($ \simeq m_U $),
$ m_{\phi_0} = 100~{\rm GeV} $, and $ m_{\phi_1} = m_{\phi_2} = 0.5 v_S $
with $ v_S = 400~{\rm GeV} $ (fig. 1), and $ 4~{\rm TeV} $ (fig. 2).
In these scatter plots, each dot corresponds to a random choice
for the set of parameters such as:
the complex phases $ \alpha_i $, $ \alpha^\prime_i $,
$ \beta $, $ \beta^\prime $ and $ \phi_S $;
the generation dependent factors
$ \kappa_i $ and $ \kappa^\prime_i $ = 2/3 -- 1;
the coupling ratio $ | \lambda_U^\prime | / | \lambda_U | $ = 0.1 -- 10;
and the Higgs mixing matrix $ {\mbox{\sf O}} $.
It is clearly found in fig. 1 that the neutral Higgs contributions
are comparable to or smaller than the experimental bounds
even for the case of $ v_S \sim v $
with $ \epsilon_{u{\mbox{-}}U} \sim 0.3 $.
On the other hand, if $ v_S \sim 1{\rm TeV} $ or larger,
they become sufficiently below the experimental bounds,
as seen in fig. 2.
It should also be remarked that
if $ \kappa_{1,2}, \kappa^\prime_{1,2} \ll 1 $
while $ \kappa_3 , \kappa^\prime_3 \sim 1 $, namely, that
only the top quark has a significant coupling to the singlet $ U $ quark,
then these flavor nonconserving
and $ CP $ violating effects  for the light quarks become much smaller.

The modification in the CKM matrix (\ref{eqn:VCKM})
induced by the $ u $-$ U $ mixings
has been estimated by the numerical calculations for the parameter
choices taken in figs. 1 and 2.  It is roughly given as
\begin{equation}
\left( \begin{array}{c}
V_{u_{\rm L}}^\dagger V_0 - V_0 \\
\epsilon_{u_{\rm L}}^\dagger V_0 \end{array} \right)
\sim \left( \begin{array}{ccc}
10^{-4} & 10^{-4} & 10^{-5} \\
10^{-4} & 10^{-4} & 10^{-4} \\
10^{-5} & 10^{-4} & 10^{-2} \\
10^{-3} & 10^{-2} & 10^{-1}
\end{array} \right) ~.
\label{eqn:dV}
\end{equation}
In particular, for the charged gauge interactions of the $ t $
quark and the $ U $ quark,
$ | V_{tb} | = 0.95 - 1 $ and $ | V_{Ub} | = 0 - 0.2 $
are obtained.
This amount of slight modification in the electroweak gauge interactions
of the quarks provide contributions smaller than about 0.2
to the oblique parameters \cite{L-S},
which are still consistent with the experimental bounds
(see for instance p. 104 in ref. \cite{ParticleData}).

The modification in the neutral currents
$ {\cal U}^\dagger \sigma^\mu V_Z {\cal U} $
coupled to the $ Z $ boson has also been estimated as
\begin{equation}
V_Z - V_Z^{(0)}
\sim \left( \begin{array}{cccc}
10^{-11} & 10^{-9} & 10^{-7} & 10^{-6} \\
10^{-9} & 10^{-7} & 10^{-4} & 10^{-4} \\
10^{-7} & 10^{-4} & 10^{-2} & 10^{-1} \\
10^{-6} & 10^{-4} & 10^{-1} & 10^{-2}
\end{array} \right) ~,
\label{eqn:dIZ}
\end{equation}
where $ V_Z^{(0)} $ represents the usual neutral currents
in the absence of $ u $-$ U $ mixings.
(The neutral currents of $ {\cal U}^c $ is not modified.)
This should be compared to the quark couplings to the neutral Higgs fields,
which have been estimated as
\begin{equation}
\Lambda_a
\sim \left( \begin{array}{cccc}
10^{-5} & 10^{-4} & 10^{-2} & 10^{-1} \\
10^{-6} & 10^{-3} & 10^{-2} & 10^{-1} \\
10^{-6} & 10^{-4} & 10^{-1} & 10^{-1} \\
10^{-6} & 10^{-4} & 10^{-1} & 10^{-1}
\end{array} \right) ~.
\label{eqn:Lam-a-num}
\end{equation}
This indicates that the FCNC's in neutral Higgs couplings
are more important than those in $ Z $ boson couplings
in these sorts of models with the quark mass matrix
of the form given in eq.(\ref{eqn:MUmatrix})
respecting the quark mass hierarchy.

As for the case of $ d $-$ D $ mixings, the neutral Higgs contributions
to the $ K^0 $-$ {\bar K}^0 $ mixing and the NEDM should be
investigated as well.
It has actually been checked by numerical calculations
that the contributions to the $ d $ quark EDM are analogous
to those obtained for the $ u $ quark EDM
in the presence of $ u $-$ U $ mixings.
It should also be mentioned that the $ CP $ violation parameter
$ \epsilon $ for the $ K^0 $-$ {\bar K}^0 $ mixing, in particular,
can be as large as $ 10^{-3} $
for $ \epsilon_{d{\mbox{-}}D} \sim 0.1 $ and $ v_S \sim v $.
Detailed analyses will be presented elsewhere.

Finally, some comments are presented
for possible variants of the present model.
(i) The complex Higgs field $ S $ may be replaced by a real field
with $ f_U^\prime = \lambda_U^\prime = 0 $.
Then, the coupling matrix $ \lambda_U $ can be made real and diagonal
by a redefinition among the $ U_0 $ and $ U_0^c $ fields.
Even in this case, similar contributions are obtained
to $ \Delta m_D $ and $ d_u $ with more than one $ U $ quarks
and the complex Yukawa couplings $ f_U $.
(If only one $ U $ quark is present, the complex phases
in $ f_U $ are absorbed by the $ u_0^c $ fields,
resulting in the vanishing of $ d_u ( \phi ) $.)
It should, however, be mentioned that
with only one real singlet Higgs field
the $ t $-$ U $ mixing is ineffective for electroweak baryogenesis.
This is because the complex phases in the $ t $-$ U $ couplings
to the real singlet Higgs field can be eliminated by rephasing
the $ U_0 $ and $ U_0^c $ fields.
(ii) It may be considered that the singlet Higg field $ S $ is absent,
regarding $ \Delta_{u{\mbox{-}}U} $ and $ M_U $ as explicit mass terms
in eq. (\ref{eqn:MUmatrix}).  Even in this case,
the significant FCNC's are still present in the quark couplings
to the standard neutral Higgs field $ \phi_0 = h_1 $.
It should, however, be mentioned that the one-loop neutral Higgs
contribution to $ d_u $ vanishes, just as does the one-loop $ Z $ boson
contribution.
This is because the standard neutral Higgs field $ h_1 $
and the Nambu-Goldstone mode $ h_2 $ couple to the quarks
in the same way.

In conclusion, flavor nonconserving and $ CP $ violating effects
of the quark mixings have been investigated in electroweak models
incorporating singlet quarks.
It is found especially that the neutral Higgs contributions
to the neutral $ D $ meson mass difference and the NEDM
are still consistent with the present experimental bounds
even for the case where the singlet Higgs mass scale is comparable
to the electroweak scale and a significant $ t $-$ U $ mixing
is present.  This, in particular, implies that there is a good chance
for these types of models with singlet quarks
to generate a sufficient amount of baryon number asymmetry
in the electroweak phase transition.
A sufficiently strong electroweak phase transition
can be realized with the singlet Higgs field
with $ v_S \sim v $, and the asymmetries of certain quantum numbers
contributing to the baryon number chemical potential
can be produced by the $ CP $ violating $ t $-$ U $ mixing
through the bubble wall.

\bigskip
We would like to thank R. N. Mohapatra and also
F. del Aguila and J. A. Aguilar-Saavedra
for informing us of their papers,
which are relevant for the present article.

\newpage

\newpage
\begin{flushleft}
{\Large{\bf{Figure Captions}}}
\end{flushleft}

\begin{description}

\item[Fig. 1]
The neutral Higgs contributions of $ | d_u ( \phi ) | $ versus
$ \Delta m_D ( \phi ) $ are plotted for the case of $ v_S = 400~{\rm GeV} $
together with the experimental upper bounds
on the NEDM (dashed line) and on $ \Delta m_D $ (dotted line).
The relevant parameters are taken to be
$ \epsilon_{u{\mbox{-}}U} = 0.3 $, $ M_U = 300{\rm GeV} $,
$ m_{\phi_0} = 100{\rm GeV} $, and $ m_{\phi_1} = m_{\phi_2} = 0.5 v_S $.

\item[Fig. 2]
The neutral Higgs contributions of $ | d_u ( \phi ) | $ versus
$ \Delta m_D ( \phi ) $ are plotted for the case of $ v_S = 4~{\rm TeV} $.
The same values are taken for the relevant parameters as in fig. 1.

\end{description}

\newpage
\begin{figure}
\begin{center}
\vspace*{1.5cm}
\includegraphics[width=13cm,height=10cm]{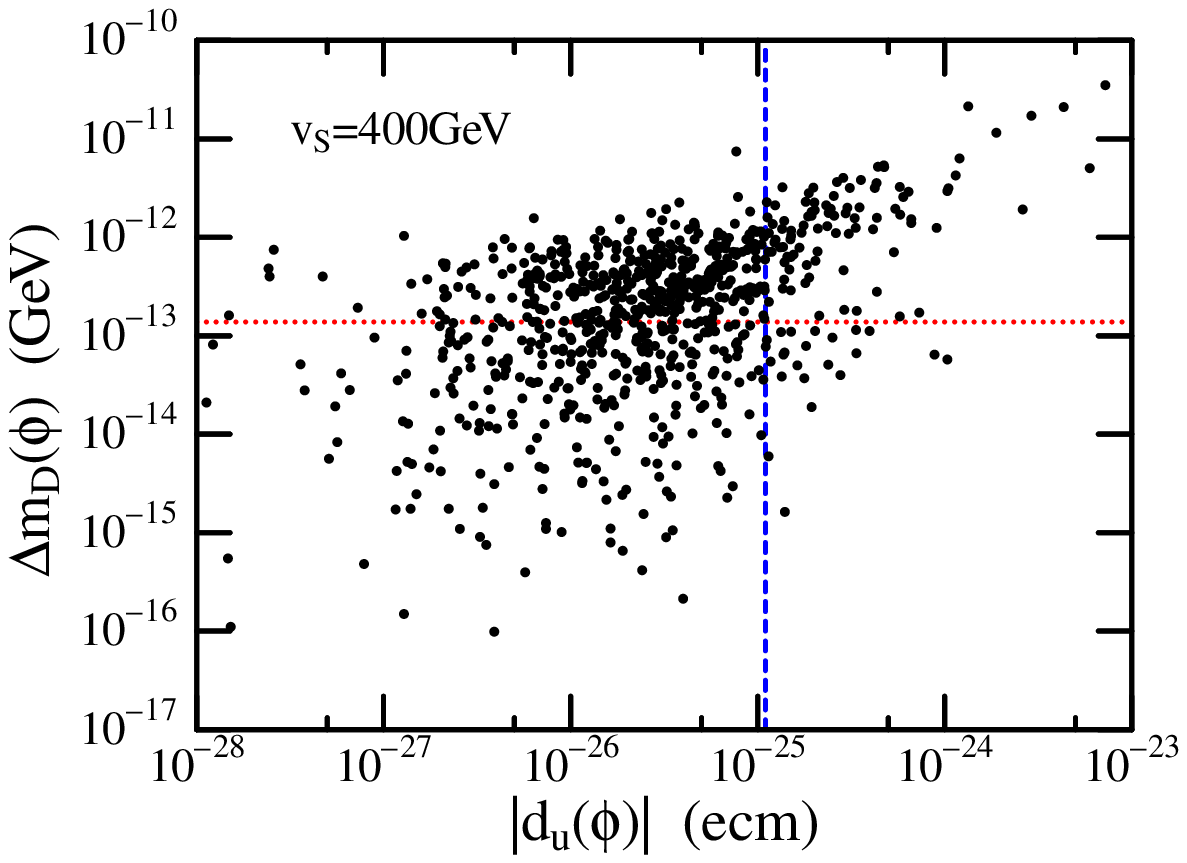}
\label{F1}
\end{center}
\end{figure}

\vspace*{3.5cm}
\begin{center}
{\Large{\bf Fig. 1}}
\end{center}

\newpage
\begin{figure}
\begin{center}
\vspace*{1.5cm}
\includegraphics[width=13cm,height=10cm]{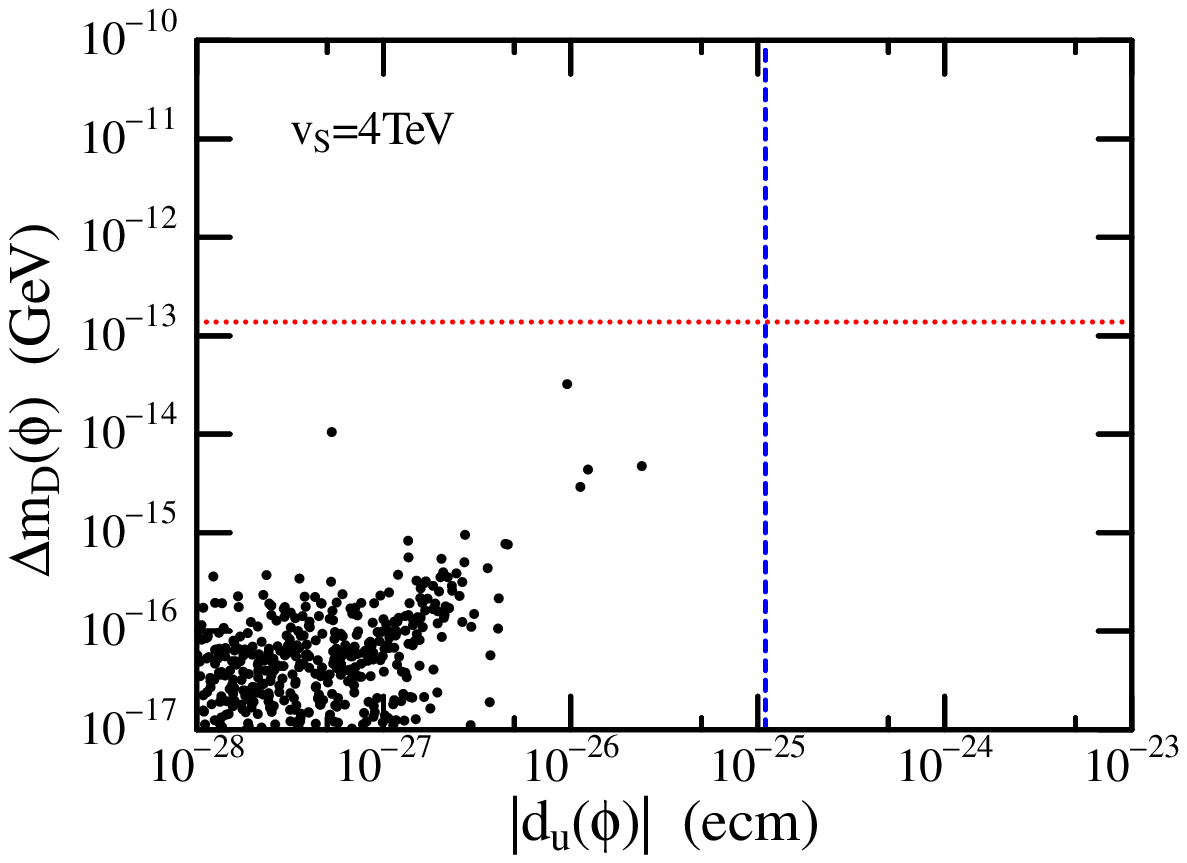}
\label{F2}
\end{center}
\end{figure}

\vspace*{3.5cm}
\begin{center}
{\Large{\bf Fig. 2}}
\end{center}

\end{document}